\documentstyle[12pt]{article}

\newcommand{\eq}{\begin{equation}}
\newcommand{\en}{\end{equation}}

\newcommand{\NP}[1]{Nucl.\ Phys.\ {\bf #1}}
\newcommand{\PL}[1]{Phys.\ Lett.\ {\bf #1}}

\newcommand{\PR}[1]{Phys.\ Rev.\ {\bf #1}}

\newcommand{\IJMP}[1]{Int.\ J.\ Mod.\ Phys.\ {\bf #1}}

\begin{document}

\hskip 12cm \vbox{\hbox{DFTT 47/91}\hbox{November 1991}}
\vskip 0.4cm
\centerline{\bf SELF-AVOIDING EFFECTIVE STRINGS }
\vskip .6 cm
\centerline{{\bf IN  LATTICE GAUGE THEORIES\footnote
{\it Work supported in part by Ministero dell'Universit\`a e della
Ricerca Scientifica e Tecnologica}}}

\vskip 1.3cm
\centerline{ M. Caselle\footnote
{email address: Decnet=(39163::CASELLE)~~Bitnet=(CASELLE@TORINO.
INFN.IT)}
and  F. Gliozzi}
\vskip .6cm
\centerline{\sl  Dipartimento di Fisica
Teorica dell'Universit\`a di Torino}
\centerline{\sl Istituto Nazionale di Fisica Nucleare,Sezione di Torino}
\centerline{\sl via P.Giuria 1, I-10125 Turin,Italy}
\vskip 3.cm
\begin{abstract}
 It is shown that the effective string recently introduced to describe
 the long distance dynamics of $3D$ gauge systems in the confining phase
has an intriguing description in terms of models of $2D$ self-avoiding
walks in the dense phase.
The deconfinement point, where the effective string becomes $N=2$
supersymmetric, may then be interpreted as the tricritical
$\Theta$ point where the polymer chain undergoes a collapse
transition. As a consequence, a universal value of the deconfinement
temperature is predicted.

\end{abstract}
\vfill
\eject

\newpage
\setcounter{page}{1}
In the string picture of the long-distance dynamics of a gauge theory in
the confining phase, the quark pairs are linked together by a thin,
fluctuating colour flux tube~\cite{string}.
 This effective string  can be described as
a conformal field theory (CFT) on a surface with quark lines as
boundaries.

Starting from  general properties of
lattice gauge theories in $D$ space-time dimensions, a specific CFT
has been recently proposed~\cite{n2} as a
model for the effective string at zero temperature and it
is identified by the following properties:
\begin{description}
  \item[i)] it is  composed, at large distances, by $D-2$
free bosons, describing the transverse displacements of the string;
  \item[ii)] each bosonic field is compactified on a circle of radius
$R_t$ related to the finite thickness of the flux tube. This allows
fermionizing  the model, which  in turn guarantees that the colour
flux tube does not self-overlap;
  \item[iii)] the boundary conditions on  quark lines are dictated
by the lattice gauge theory (LGT) and fulfil a sort of orthogonality
constraint such that, whenever the quark and antiquark lines coincide,
 the effective string vanishes . This   forces the adimensional radius
$\nu=\sqrt{\pi\sigma}R_t$  to be exactly $\nu=1/4$.
\end{description}

 This model is
in good agreement with results of Montecarlo simulations both in
3~\cite{n1} and in 4~\cite{n0} dimensions. In the following we will
concentrate in particular on three dimensional LGT.
In this case one can also argue some interesting information on the
effective string picture at the deconfinement point, which turns out to
be related with a twisted $N=2$ supersymmetric CFT's with
$c=1~$\cite{gc}.

In $D=3$ the model of ref~\cite{n2} is simply a particular case of the
gaussian model with action
\eq
S=\frac{g}{4\pi}\int  d\tau d\varsigma\,\partial\phi\partial\phi.
\label {action}
\en
Since the field $\phi$ is compactified on a circle, topologically non
trivial configurations are allowed. The partition function on the
torus is obtained by taking into account the quantum
fluctuations around each one of the classical solutions. A standard
calculation gives (in the notation of ref~\cite{fsz1,fsz2}):
\eq
Z_c[g,1]=\frac{1}{\eta(q)\eta(\bar q)}\sum_{n,m\in\bf{Z}}
q^{(\frac{n}{\sqrt{g}}+m\sqrt{g})^2/4}
\bar q^{(\frac{n}{\sqrt{g}}-m\sqrt{g})^2/4}.
\label{zg1}
\en
More general partition functions can be written if, for any reason,
the solitonic configurations are somehow constrained:
\eq
Z_c[g,f]=\frac{1}{\eta(q)\eta(\bar q)}\sum_{n\in{\bf Z}/f,m\in{\bf Z}f}
q^{(\frac{n}{\sqrt{g}}+m\sqrt{g})^2/4}
\bar q^{(\frac{n}{\sqrt{g}}-m\sqrt{g})^2/4}.
\label{zgf}
\en
It is easy to show that
$$Z_c[g,f]=Z_c[gf^2,1].$$

It is well known  that  the ordinary  Dirac fermion is exactly
described by ~~$Z_c[g=1/2,1]$~.  Other Dirac fermions with non trivial
spin structures are described by $f\not= 1$. The model of ref~\cite{n2} is
given by $g=1/2$ and $f=1/2$. The spectrum of conformal weights
is given by
\eq
h_n=\frac{n^2}{32}~~,~~n=0,1,\dots
\label{sb}
\en

If we replace the torus with an infinite strip of width $R$, which is
the appropriate world sheet of the conformal theory describing the
effective string, the spectrum of physical states propagating along the
strip  is a subset of those listed in eq . (\ref{sb}),  depending on the
boundary conditions on either side of the strip. According to the
point $iii)$ , the gauge theory fixes uniquely
these boundary conditions \cite{n2}  in such a way  that the only
propagating, physical states belong to the  twisted sector of a free
fermion with boundary phase $1/4$, corresponding to a ground state of
conformal weight $h=1/32$. As a consequence, the static potential
between the two quarks  is given by~\cite{n2}
\eq
V(R)=\sigma R \,+\,k\,+\,\tilde c \frac{\pi}{24R}\,+\,O(R)~~;
{}~~\tilde c=c-24h=1/4~,
\label{pot}
\en
\noindent
where $\sigma$ is the string tension , $k$ is a non- universal constant,
$\tilde c$ is the {\em effective } central charge \cite{ceff}
controlling the universal Casimir energy generated by the finite size of
the strip \cite{blote}.

\vskip .3cm
The constraint $f=1/2\,$, which is a direct consequence of the point
$iii)$ , has also a nice interpretation in
the context of the critical $O(n)$ models (or, more precisely, in terms
of the corresponding SOS model~\cite{baxter}). This model can be defined
through analytic continuation in $n$. The original $O(n)$ model is
defined initially for $n\in N$, $n\geq2$ by
\eq
Z_{O(n)}=\int \prod_i d\vec S_i\hskip 0.2cm \prod_{<j,k>}(1+\frac{1
}{T}\vec S_j\vec S_k)~~~,
\en
where $\vec S$ is $n$-component vector such that $|\vec S|^2=n$. By high
temperature expansion one gets on the hexagonal lattice:
\eq
Z_{O(n)}=\sum_{graphs}(\frac{1}{T})^{N_b}n^{N_l}~~~.
\label{zon}
\en
The sum is over all configurations of non intersecting, self- avoiding
graphs, $N_l$ is the number of loops and $N_b$ is the total number of
bonds in the graph. One can analytically continue (\ref{zon}) to
$n\in{\bf R}$. The model can then be explicitly transformed into a SOS
model (see ref.~\cite{nienhuis} for details) and in this way can be
solved. It can be shown that the model is critical for $n\in[-2,2]$, and
in the continuum critical limit, renormalizes onto the gaussian
model defined by the
action (\ref{action}) and with a coupling constant related to the
parameter $n$ by
\eq
n=-2\cos \pi g.
\label{g}
\en
The two branches of the arc cosine with $g\leq 2$ have well defined
different meanings.
Indeed it is known that the phase diagram of the model has a  rich
structure: it is critical for ${\cal {T}}_c=\sqrt{2+\sqrt{2-n}}$ ; this
is the so called dilute phase and it is related to the branch
$g\in[1,2]$ . But this model is also critical in the whole region
${\cal{T}}<{\cal{T}}_c$ (the dense
phase),  corresponding to the other branch $ g\in[0,1]$ .
The global properties of the system can be investigated by looking for
instance at the theory formulated on the cylinder or on the
torus~\cite{fsz1,fsz2}.
Due to the particular mapping  into an SOS model {\sl also defects
at the boundaries of the type $\phi=\pi$} (and not only $\phi=2\pi$)
are allowed. This is coded in eq.(\ref{zgf}) by $f=1/2$.

\vskip .3 cm
Among all the possible values of $n$ in (\ref{zon}), a special role
is played by the limit $n\to 0$ , which is related to the self avoiding
walks (SAW) problem. Looking at (\ref{g}) we see that this exactly
corresponds to our choice $g=1/2$.
This seems to indicate that the model
proposed in ref~\cite{n2} gives an effective description of the CFT in
the surface bordered by the quarks
 in terms of self-avoiding walks {\sl in the dense phase}.

The CFT associated to the SAW problem in the dense phase is non unitary,
with central charge $c=-2$  . The spectrum of conformal weights is
(see for instance~\cite{saw}):
\eq
h_n=\frac{n^2-4}{32}~~,~~n=0,1,\dots
\label{pesi}
\en
\noindent
which looks different from that of the gaussian model given in eq.
(\ref{sb}).
Actually  the partition function of this non unitary theory on the torus
 and on the cylinder may be expanded also in terms of $c=1$ characters
\cite{fsz1}\footnote{The central charge $c=-2$ and the non unitary
spectrum of eq.(\ref{pesi}) can be obtained from that of the gaussian, $c
=1$ model of eq.(\ref{sb}) adding a suitable charge at infinity, such
that topologically non trivial loops have the same weights of the
trivial ones in the partition function} .
In particular, the interquark potential is a function of
the effective central
charge $\tilde c$
\eq
\tilde c=c-24h_l= 1-\frac{3l^2}{4}~,
\label{csaw}
\en
\noindent
where $l$ labels the lowest physical state which can propagate along the
strip. If $l=1$, we get exactly $\tilde c=1/4$ as in eq.(\ref{pot});
this may be understood as follows.

\vskip .3 cm
Notice that the equivalence between SOS model and the standard SAW
problem is valid only on  a simply connected surface. On a cylinder
with quark lines as boundaries we must allow an odd number of
self-avoiding loops wrapping around the cylinder.
In fact, only in this way
the two opposite boundaries of the cylinder have a shift in the boson
field $\phi(R)-\phi(0)=\pi$ , so that the boundary conditions fulfil
the orthogonality constraint described in the point $iii)$.
As a consequence, the ground state ( corresponding to $l=0$ ) is
projected out , so  we have  $l=1$ in eq.(\ref{csaw}) , in agreement
with eq.(\ref{pot}). In such a case
the partition function can be expanded in terms of fermionic
determinants with boundary phases $\nu=m/4$ with $m=0,1,2,3$ as shown
in ref.\cite{n2,saleur}.

\vskip.3 cm

We want to give an heuristic argument suggesting that there is a set
surfaces, spanned by the colour flux
tubes in 2+1 dimensions, which generate 2D SAW configurations. In
particular the rough surfaces, which control the functional integral in
the region where the string picture works, will generate SAW in the
dense phase. We shall use this argument to study the behaviour of the
effective string as a function of the physical temperature of the gauge
lattice system.

\vskip .3 cm
Consider a  pair of quark -antiquark lines parallel to the imaginary
time axis of a $3D$ cubic lattice of size $\infty^2\times L$ , with
periodic boundary conditions in the time direction, describing a gauge
system at the temperature $T=1/L$ .
The surface swept by the elementary colour flux tube joining this pair
of quarks in the strong coupling region is topologically a cylinder ,
with the two quark lines of length $L$ as boundaries.
Describing this surface as the world sheet of a free bosonic string is
too drastic an approximation, which does not allow to take into
account the correct boundary conditions~\cite{n2}.

\vskip .3 cm
A better picture of the observed flux tube , which has a finite
thickness, is obtained surrounding the quark
sources by a set of flux rings, like in  fig. 1. In the 3D lattice
these rings sweep toroidal surfaces linked to the quark lines.

\begin{center}
\begin{picture}(240,80)(0,0)

\put(70,40){\circle*{3}}
\put(70,40){\line(1,0){10}}
\put(80,40){\line(0,1){20}}
\put(80,60){\line(1,0){10}}
\put(90,60){\line(0,-1){10}}
\put(90,50){\line(1,0){20}}
\put(110,50){\line(0,1){10}}
\put(110,60){\line(1,0){25}}
\put(135,60){\line(0,-1){40}}
\put(135,20){\line(1,0){20}}
\put(155,20){\line(0,1){20}}
\put(155,40){\line(1,0){25}}
\put(90,40){\oval(80,30)}
\put(80,40){\oval(44,10)}
\put(180,40){\circle*{3}}
\put(160,40){\oval(80,20)}
\end{picture}
\end{center}
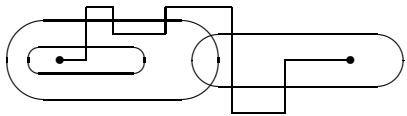
\centerline{\it fig. 1}
{\it A colour flux tube joining a quark pair ( the two dots ),
surrounded by flux rings linked to the quark sources.}
\vskip .2 cm

The intersections of the cylinder with the tori are self-avoiding loops
which can wrap the tori (there is an odd number of homotopically non
trivial SAW for each torus).  When these surfaces are all smooth,
it is immediate to see, using strong coupling arguments, that
their contribution to the correlation function of the two quark lines is
partly suppressed by the vacuum diagrams (in $Z_2$ gauge theory the
cancellation is complete). On the contrary, whenever one of these
surfaces is crumpled up , there is a strong contribution due to excluded
volume effects .

\vskip .3 cm
There is a sort of duality between the configurations
of the cylinder and those of the tori: If the cylinder is very rough,
the tori are forced to be smooth by excluded volume effects, and
vice versa. The former configurations  are those
 that should dominate at very low temperature and
large distances, where the configurations available for the cylinder
grow  with the area between the two quark lines, while the tori
configurations grow only with the length $L=1/T$ of the quark lines.
In such a case it is reasonable to assume that the
intersections of the  rough cylinder with the tori are self-avoiding
loops in the dense phase. Probably the above considerations might be made
more precise introducing, for instance, the concept of winding of SAW
around the (quark) sources , like in ref~\cite{winding}. We use them
only to guide our intuition in the study of the confining phase.

\vskip .3 cm
Let us push now a bit further our construction. As we increase the
physical temperature $T$ of the LGT, the tori become more and more rough
( indeed the tori configurations will dominate the liberated quark
phase). Then, the excluded volume effects become more and more important
, and produce vacancies in the dense phase. These vacancies simulate the
introduction of a solvent in a polymer chain . At a
certain value of the concentration and of the temperature $\Theta$,
the polymer undergoes a collapse transition, which corresponds to a
tricritical point~\cite{degennes}.

\vskip .3 cm
It is now tempting to argue that this  transition coincides
precisely with the quark deconfinement. Actually the partition function
for the model of SAW at the tricritical $\Theta$ point has been
calculated in ref.\cite{theta,saleur} and turns out to be described by
a gaussian model (\ref{zgf}) with $g=2/3$ and $f=1/2$. Remarkably enough
, this CFT coincides exactly with the model of the effective string at
the deconfining temperature proposed in ref.\cite{gc} .

\vskip .3cm
Notice that this model was based on a
completely different argument, namely that at the deconfinement point ,
where the effective string disappears, also the Casimir effect must
vanish, implying $\tilde c=0$. This may be also understood by
noting that $\tilde c$ measures the number of physical degrees of
freedom of the effective string, so it must go to zero when the string
vanishes. The fact that the effective string at the deconfining
transition has no (local) physical degrees of freedom suggests that it
might be described by a topological conformal theory . Indeed it has
been shown ~\cite{gc} that there is  only one
CFT with $c=1$ and vanishing effective central charge
and it corresponds to the compactification radii
\eq
r\equiv \sqrt{\frac{gf^2}{2}}=\frac{n}{2\sqrt{3}}~, ~ (n=1,2,3,6)~,
\label{radii}
\en
where the $c=1$
conformal symmetry is promoted to a $N=2$ supersymmetry \cite{susy}
which supports the topological nature of the model.

\vskip .3 cm
On the other  hand, these compactification radii are simply related to
the physical temperature $T$ of the $3D$ gauge  lattice theory
through $r=\sqrt{\frac{\sigma}{\pi}}/2T$ ; for this reason they have
been used \cite{gc} to select four special values of $T$, one of which
should correspond to the deconfining transition $T_c$.

\vskip.3cm
Now, the analogy with the tricritical point of the polymer chain we have
established in the present paper  allows us to identify the correct
value of the transition point. According to eq.(\ref{radii}),  the
value of $g$ and $f$ singled out by the $\Theta$ point corresponds to
$n=1$. It turns out that the corresponding value of the critical
temperature $T_c/\sqrt{\sigma}=\sqrt{3}/\sqrt{\pi}$ is universal and
fits well to the numerical data  \cite{gc}.

\vskip 1.5cm
We thank R. Fiore, I. Pesando, P. Provero and S. Vinti for many useful
and interesting discussions.

\end{document}